\documentclass[structabstract]{aa}
\usepackage{graphicx}
\usepackage{txfonts}
\usepackage{natbib}
\def\simless{\mathbin{\lower 3pt\hbox
     {$\rlap{\raise 5pt\hbox{$\char'074$}}\mathchar"7218$}}}   
\def\simmore{\mathbin{\lower 3pt\hbox
     {$\rlap{\raise 5pt\hbox{$\char'076$}}\mathchar"7218$}}}   

\begin{document}
\title{
A quantitative explanation of the type-B QPOs in GX 339--4
}

\subtitle{}

\author{
Nikolaos D. Kylafis\inst{1,2} 
Pablo Reig\inst{2,1}
\and
Iossif Papadakis\inst{1,2}
}

\institute{
University of Crete, Physics Department \& Institute of Theoretical and
Computational Physics, 70013 Herakleion, Crete, Greece\
\and
Institute of Astrophysics, Foundation for Research and Technology-Hellas
(FORTH-IA), 71110 Heraklion, Crete, Greece
}
\date {Received ; Accepted ;}


\abstract
{
Type-B quasi periodic oscillations (QPOs) in black-hole X-ray binaries
(BHXRBs) are a class of low-frequency QPOs that are observed in the soft intermediate 
state in the rising and the declining phases of an outburst.
They are suspected to result from the precession of the jet that is
ejected from the source.
}
{
The objective of the present work is to investigate in detail the 
emissivity of the jet in hard X-rays and to see whether the type-B QPOs from
GX 339-4, which is the best studied black-hole transient, can be explained
quantitatively with a precessing jet.
}
{
We used our simple jet model, which invokes Comptonization in the jet,
and examined the angular dependence of the upscattered photons that emerge
from the jet and their energy distribution, which is a power law.  
}
{
Due to the  elongation of the jet, assisted by the bulk 
motion of the electrons, the angular distribution 
of the emerging hard X-ray photons from the jet is not isotropic.  
More importantly, the photon-number spectral index, $\Gamma,$ is 
an increasing function of the polar angle, $\theta,$ with respect to the
axis of the jet.  
If the jet is fixed, then an observer at infinity sees the photon
index, $\Gamma,$ which corresponds to this specific
observational direction.  However,
if the jet is precessing, then the observer sees a periodic variation of
$\Gamma$ with the precession period.
Such a periodic variation of $\Gamma$ has been observed in GX 339-4
and in this work, we reproduce it quantitatively, using our model.
}
{
Our jet model nicely explains through quantitative means 
the type-B QPOs seen in GX 339-4 as originating from a precessing jet.
The given model has previously explained several observed correlations thus far. }


\authorrunning{Kylafis, Reig, \& Papadakis 2020}

\titlerunning{Type-B QPOs in GX 339-4}

\maketitle


\section{Introduction}

During their outbursts, black-hole X-ray binaries (BHXRBs) exhibit  a
characteristic hysteresis curve in a hardness-luminosity diagram 
\citep{miyamoto95,homan01,belloni05,homan05,gierlinski06,remillard06,fender09,motta09,belloni10,munoz-darias11a,stiele11}.  
As the luminosity increases during an outburst, the sources exhibit a range of
characteristic states, which, in the classification of \citet{belloni05}, are
referred to as: quiescent, hard, hard-intermediate, soft-intermediate, and
soft.   As the luminosity decreases over the decline of an outburst, the states
are traced back in the opposite order.

During the rise of an outburst, a compact jet is always seen in the first three
states. By the soft intermediate state, a jet had previously been suspected to
be present  but it is confirmed that it has been observed \citep{russell20}.  
In the soft state, the jet disappears.  

As the luminosity decreases, the jet is suspected to have been re-established in
the soft intermediate state, while it is detected  in the hard intermediate
state \citep{corbel13}.  Near the end of  the hard intermediate state, the jet
becomes compact and stays so  in the hard and the quiescent states.

These BHXRBs exhibit three types of low-frequency quasi periodic oscillations
(QPOs), A, B, and C \citep{wijnands99,casella05,motta12}. Here, we concentrate
on the type-B QPOs, which are seen only in the soft intermediate state.  The
characteristic frequency of the type-B QPOs is in the  relatively narrow range
of 1 - 6 Hz \citep{motta11}, while at high-flux intervals, the range becomes
even narrower (4 - 6 Hz) \citep{casella04,motta11}.

\citet{kylafis15} speculated that Type-B QPOs are associated  with the last
"gasps" of the jet in the rising part of an outburst and with the
re-establishment of the jet in the declining part.   This idea was supported by
the observational fact that  the type-B QPOs are stronger in low-inclination
sources \citep{motta15}.

\citet{stevens16} performed phase-resolved  spectroscopy of the type-B QPOs in
GX 339-4.  They found that the photon-number spectral index $\Gamma$ varies
sinusoidally  from $\sim 2.3$ to $\sim 2.6$ with the QPO frequency, $\nu = 5.2$
Hz.  They   suggested that these variations can be explained with a precessing
jet.

It is generally assumed that the hard X-ray power law with  photon index
$\Gamma$ in BHXRBs is produced by Compton upscattering of soft photons. The soft
photons come from the thin disk \cite[][hereafter SS-disk]{shakura73} and they
are  Comptonized in the corona \citep{esin97,done07}. The corona is naturally
taken to be the hot inner flow, that is, the flow inside the SS-disk, which is
geometrically thick and optically thin \citep{narayan94,narayan95}.  

This picture, however, neglects the fact that the jet is fed from the hot
inner flow and there is no boundary between the two.  
Thus, soft photons from the SS-disk
may initially be scattered in the corona, but most of them cannot
escape without entering the jet and being scattered there.  Since
photons forget their past history after a few scatterings, it is the
Comptonization in the jet that determines the observed $\Gamma$
\citep{reig03,giannios04,giannios05,kylafis08,reig15,kylafis18,reig18}.

In \citet{reig19}, we showed that the observed photon index $\Gamma$ depends on
the inclination angle of  the source.  If we could see the same source from
different directions,  the spectra would become softer  (i.e., with larger
$\Gamma$) as the inclination angle increases  \citep[see Fig. 6 in][]{reig19}.
But this is exactly what would also happen for a source, with a fixed  viewing
angle, if the hard X-rays come from the jet and the jet is  precessing. For this
reason, we examine in this letter how  $\Gamma$ depends on the viewing angle and
we quantitatively explain the  periodic variation of $\Gamma$ found by
{\citet{stevens16}.  In \S\ 2 we briefly discuss our model, in \S\ 3 we present
the results of  our calculations, in \S\ 4 we discuss our work, and in \S\ 5 we
give our  conclusions.

\section{Jet model}

For our calculations, we used a simple jet model, which was previously
presented  in \citet{reig19}.  Here we describe it briefly.

The model assumes a parabolic jet, 
as the observations suggest \citep{asada12,kovalev20},
with an acceleration zone near its base
and constant flow beyond it.  The parabolic 
shape of the jet implies that the
radius of the jet as a function of distance from the black hole is 
given by $R(z)=R_0 (z/z_0)^{1/2}$, where $z_0$ and $R_0$ denote 
the distance of the base of 
the jet from the black hole and the radius of the jet, respectively.  
As in our previous work, the acceleration zone is taken to be between
$z_0=5R_g$ and $z_1=50R_g$, where $R_g=GM/c^2$ is the gravitational radius.
The flow speed in the jet is again taken to be
$v_{\parallel}(z)=(z/z_1)^{1/2}v_0$, for $z \le z_1$ , and 
$v_{\parallel}(z)=v_0=0.8 c$ for $z>z_1$.
The electron density $n_e(z)$
in the jet is determined from the continuity equation
and for $z>z_1$ , it falls to $n_e(z) \sim 1/z$.
The distribution of the electrons in Lorentz $\gamma$ 
in the jet is assumed to be a steep
power law, thus, most of the electrons have $\gamma = \gamma_{\rm min} =
1/\sqrt{1-(v_{\parallel}^2 +v_{\perp}^2)}$, where $v_{\perp}=0.4c$ is the
perpendicular component of the electron velocity in the jet.

We parametrize our models either with the Thomson optical depth, 
$\tau_{\parallel}$ , along the axis of the jet 
or with the radius $R_0$ at the base of the jet.  
The rest of the parameters are kept at the constant values given above.
Since the jet
in the soft intermediate state is very weak, the values of $\tau_{\parallel}$
are relatively small.  The perpendicular Thomson optical depth 
in the jet at height, $z,$ is 
$\tau_{\perp}(z)= n_e(z) \sigma_T R(z)$, where $n_e(z)$ is the 
electron number density at a height, $z,$ in the jet, which can be determined 
from the continuity equation if $\tau_{\parallel}$ is given.

Soft blackbody photons ($kT_{bb}=0.2$ keV) from the SS-disk enter  the jet at
its base and their random walk in the jet is followed by a Monte Carlo code. 
The model and the code have been used for many years with many successes.  In
particular, the model explains the spectra of BHXRBs from radio to hard X-rays
\citep{giannios05}, the time lag versus Fourier frequency \citep{reig03}, the
narrowing of the autocorrelation function with photon energy \citep{giannios04},
the correlation between time lag and photon index $\Gamma$
\citep{kylafis08,kylafis18,reig18}, the correlation between cutoff energy
and phase lag \citep{reig15}, and the inclination dependence of the
correlation between time lag and $\Gamma$ \cite{reig19}.

\begin{figure}
\centering
\includegraphics[angle=0,width=8cm]{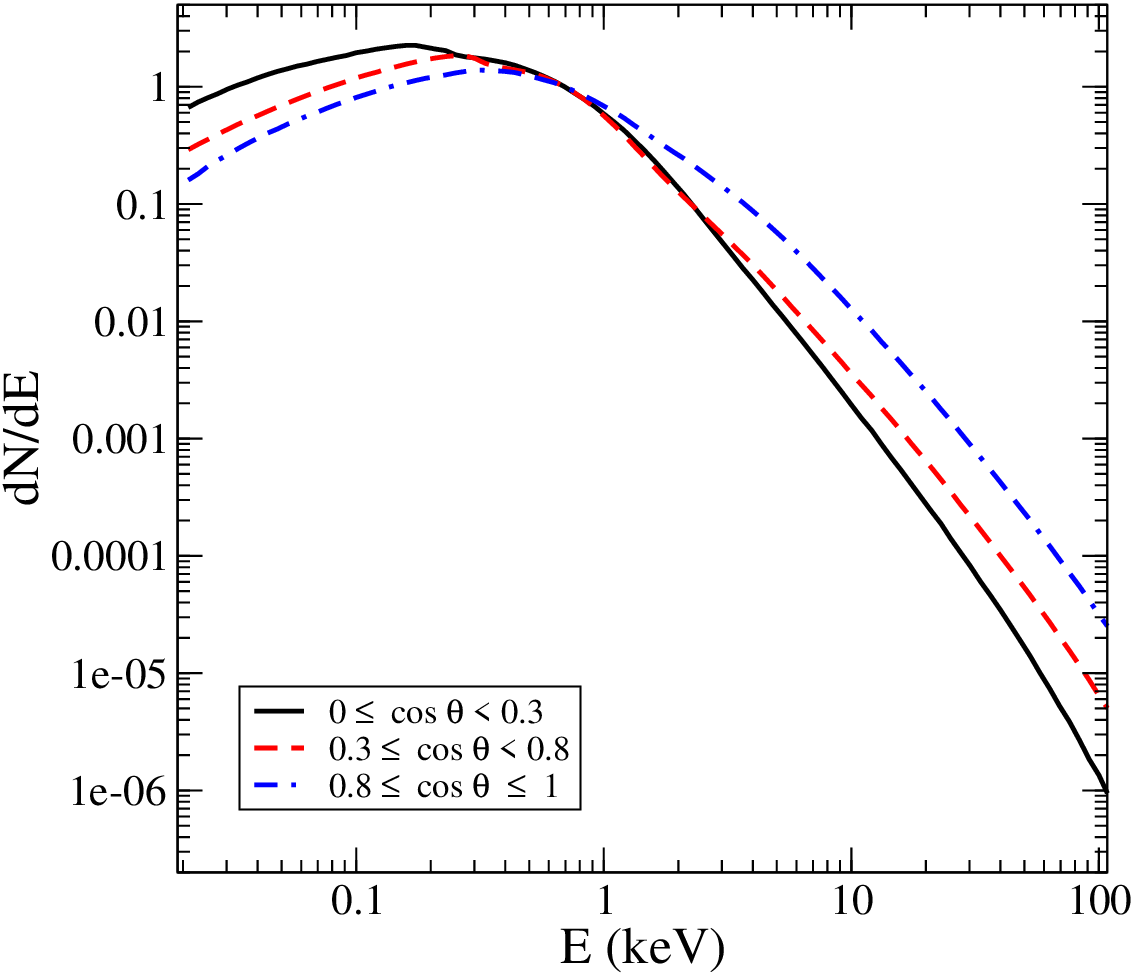}
\caption{
Emergent spectrum from the jet in three direction bins. 
}
\label{Fig1}
\end{figure}

\section{Results}

We performed Monte Carlo calculations for the radiative transfer
of blackbody photons in the jet and computed the
emergent spectra as a function of the angle $\theta$ between the
line of sight and the jet axis.

The spectra are power laws with photon index $\Gamma$ and 
have a high-energy cutoff.  
In Fig. 1, we show a representative example of
the emergent spectra from the jet 
($\tau_{\parallel} = 2.5, R_0=100R_g$),
in three direction bins:  
$0.8 < \cos \theta < 1.0$ (dot-dashed line), 
$0.3 < \cos \theta < 0.8$ (dashed line), and
$0.0 < \cos \theta < 0.3$ (solid line).
We have normalized the spectra by the flux at 0.7 keV to emphasize
the change in the slope at high energies.
It is clear from this figure
that the photon index $\Gamma$ of the power law depends
on the direction of observation.

In Fig. 2, we show model results of $\Gamma$ versus
$\cos \theta$ for various values of $R_0$ and a fixed value of
$\tau_{\parallel}= 2.5$.
We see more quantitatively than in Fig. 1 
that the photon index $\Gamma$ has a strong dependence on the viewing
angle $\theta$.  The spectra have $\Gamma$s that range from 
$\sim 2.2$ to $\sim 3.1$,
consistent with the observations in the soft intermediate state.  For
low values of $\theta$ (i.e., for sources seen face-on), the spectra are
relatively hard ($2.2 \simless \Gamma \simless 2.3$).
For values of $\theta$ near $\pi/2$, the spectra are 
significantly softer and $\Gamma$ ranges from $\sim 2.6$ to $\sim 3.1$.
This requires a qualitative explanation.

Let us consider a "static" Comptonizing spherical cloud of optical  depth of
$\tau,$ equaling a few units, and seed photons coming in from   outside, along
the diameter.  The first scattering occurs, on average, one mean-free path into
the cloud, that is, optical depth equal to one from the surface. There, they
scatter isotropically, but after that,  the photons see an optical depth of
$2\tau-1$ in the forward direction (the direction of entrance), $\sim 1$ in the
backward direction, and intermediate optical depths at intermediate directions.
Given the velocity distribution of the electrons, the photon index  produced by
Comptonization depends only on the optical depth. Therefore, photons that emerge
in the forward direction would have scattered  more times than those in the
backward direction.  As a result, the spectra are harder (smaller $\Gamma$) in
the forward direction and they soften monotonically from the forward to the
backward direction. Thus, even if the jet had no bulk velocity, the emergent
spectra would be  harder in the forward direction because the seed photons enter
at the base of the jet. This effect is enhanced if there is significant bulk
motion because the  photons are ``pushed'' in the forward direction by the
flow.  We have found, based on our jet model, that the photons that have
scattered the largest number of times in the jet come out preferentially in the
forward direction.

Another result shown in Fig. 2 is that 
as the radius $R_0$ increases, the spectra become harder (smaller
$\Gamma$).  This is because the perpendicular optical depth $\tau_{\perp}$
increases and as a result, the soft photons are trapped in the jet
and scatter more times.

\begin{figure}
\centering
\includegraphics[angle=0,width=8cm]{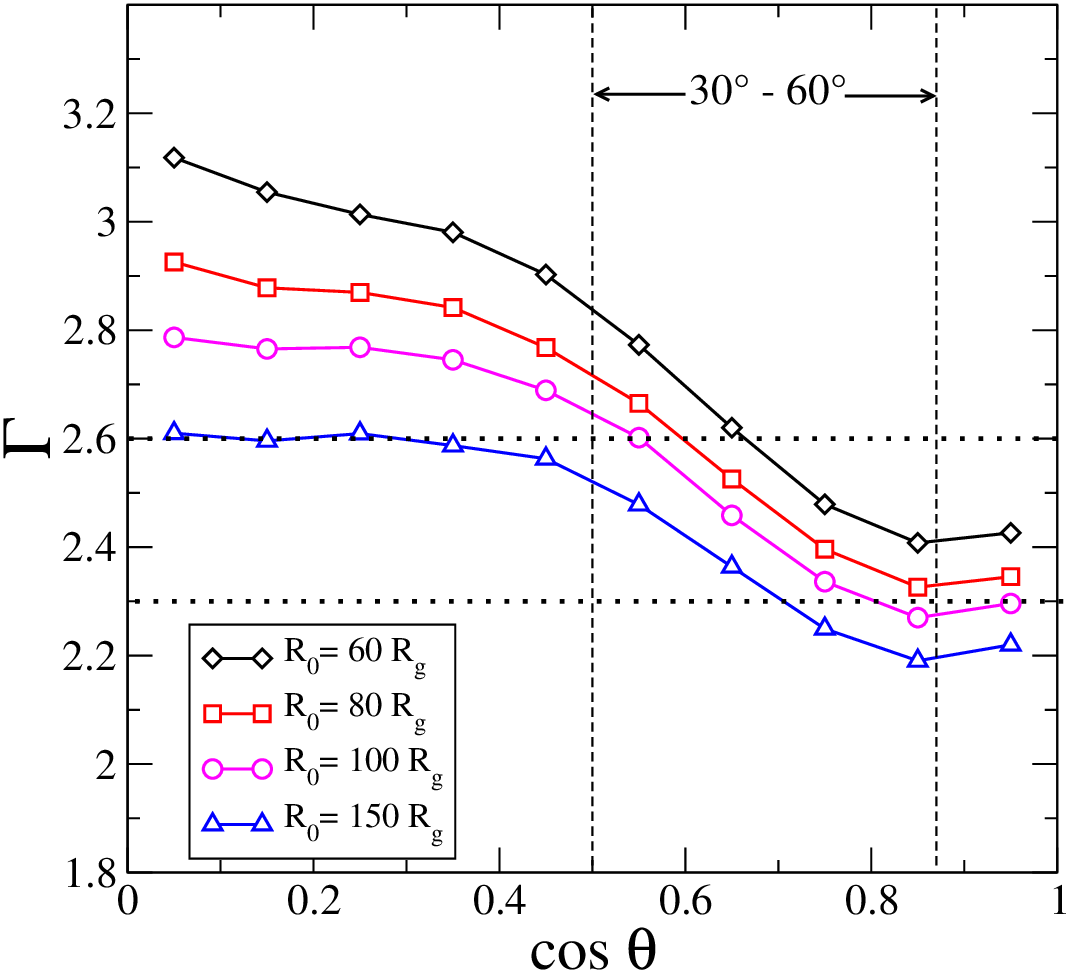}
\caption{
Model results for the relation between the photon index $\Gamma$
and the observation angle $\theta$ for $\tau_{\parallel}= 2.5$ and
various values of $R_0$.
The two horizontal dotted lines place bounds on the range $2.3 < \Gamma < 2.6$
and the two vertical dotted lines place bounds on the range $45 \pm 15$ degrees.
High-inclination systems are on the left and low-inclination ones
are on the right
}
\label{Fig2}
\end{figure}

The two horizontal dotted lines give the range of $\Gamma$ as seen by
\citet{stevens16} in GX 339-4 in the periodic variation of $\Gamma$ with the
type-B QPO frequency.  The inclination angle of GX 339-4 is believed to be $\sim
45$ degrees \citep{shidatsu11,furst15}.  Thus, the two vertical dotted lines
allow for a precession angle of 15 degrees.  The bounds in $\Gamma$ and $\cos
\theta$ place significant constraints on the models.  For $\tau_{\parallel}=
2.5$ and the fixed values of the other parameters discussed in \S\ 2, 
precessing jets with $R_0 \approx 100 R_g$ may explain  the periodic variation
of $\Gamma$ with the type-B QPO frequency.

In Fig. 3, we show model results of $\Gamma$ versus $\cos \theta$ for
various values of $\tau_{\parallel}$ and $R_0=100 R_g$.
The curves have the same morphology as in Fig. 2.  As $\tau_{\parallel}$
increases, so does $\tau_{\perp}$, and the spectra become harder.
Not surprisingly, the bounds in $\Gamma$ and $\cos \theta$ select
models close to the ones selected in Fig. 2, namely $R_0 = 100 R_g$ 
and $\tau_{\parallel} \approx 2.5$.

\begin{figure}
\centering
\includegraphics[angle=0,width=8cm]{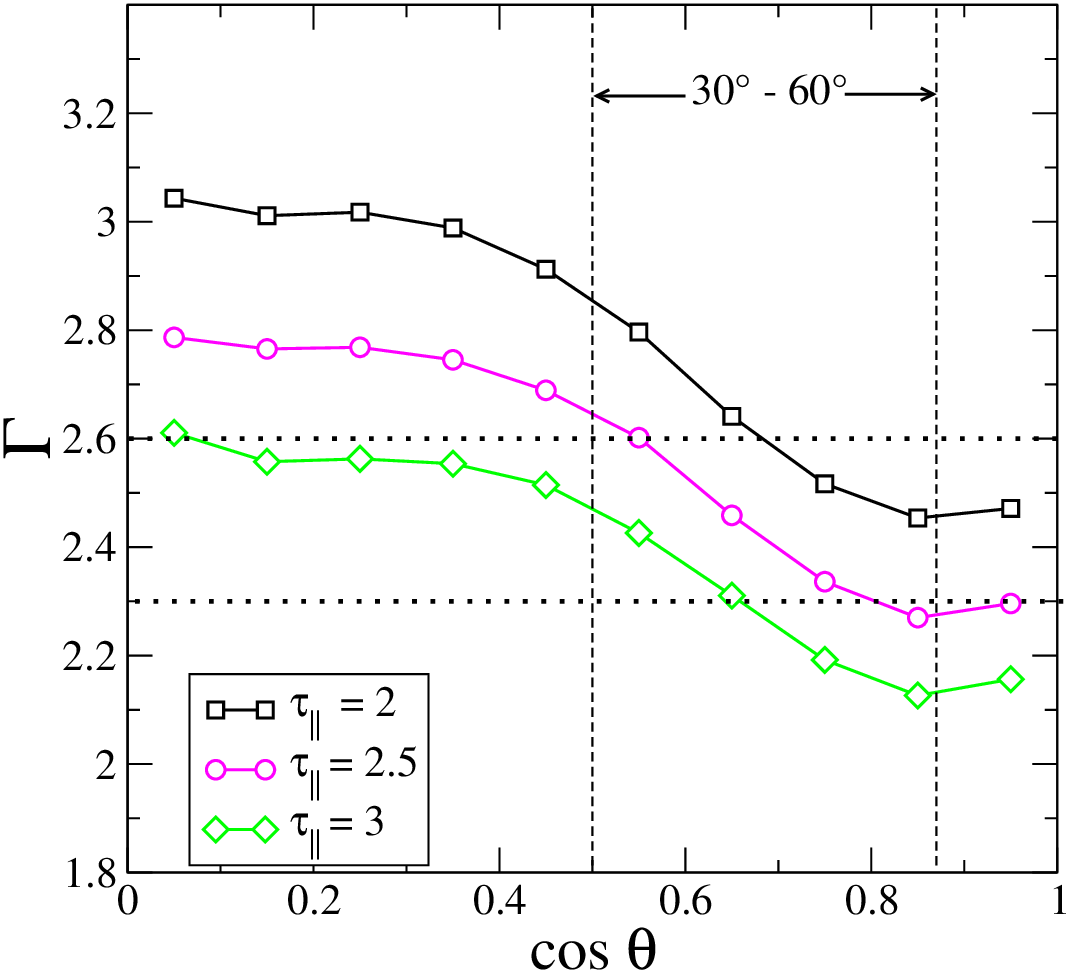}
\caption{
Model results for the relation between the photon index $\Gamma$
and the observation angle $\theta$ for $R_0= 100 R_g$ , and
various values of $\tau_{\parallel}$.
The horizontal and vertical dotted lines have the same significance as in Fig. 2.
}
\label{Fig3}
\end{figure}

\section{Discussion}

Their X-ray spectra and time variability are the two main observational quantities
in BHXRBs. A lot of effort has been put into explaining these quantities
and building models that reproduce them satisfactorily.

The spectra 
alone do not place significant constraints on the models, either on
the geometry or on the type of electrons, whether thermal or 
non-thermal.  Given enough freedom on the geometry, the optical
depth, and the distribution of electron velocities, practically
any observed X-ray spectrum can be reproduced.  In other words, 
the X-ray spectra of BHXRBs are infinitely degenerate and, by 
themselves, they cannot uniquely constrain any physical parameters.

On the contrary, time variability properties (e.g., the properties 
of QPOs, time lags, power-spectra, etc.) can place significant constrains 
on theoretical models. One such time-variability property is the one 
we study in this work, which was first published by \citet{stevens16}. 
They performed
phase-resolved spectroscopy on the type-B QPOs in GX 339-4 and they 
showed that the shape of the hard X-ray spectrum varies with the
QPO phase.  In particular, the photon-number spectral index
$\Gamma$ varies sinusoidally with the QPO frequency.  
Rather intuitively, \citet{stevens16} proposed that this 
variation of $\Gamma$ is due to a precessing jet.

The above proposal, along with the fact that Comptonization in a jet results in
direction-dependent energy spectra \citep{reig03,reig19}, have led us to a
quantitative investigation of  the dependence of $\Gamma$ with the QPO phase in
the range of $2.3 \le \Gamma \le 2.6$. For an inclination angle of $\theta = 45$
degrees for GX 339-4, we find that precession of the jet, resulting in
observational direction variation in the range of $30 < \theta < 60$ degrees,
can  quantitatively explain type-B QPOs.

We searched the parameter space for $R_0$ and $\tau_{\parallel}$
and found that if $R_0 \approx 100 R_g$ and 
$\tau_{\parallel} \approx 2.5$ then a precessing jet can explain the variation of $\Gamma$
observed by \citet{stevens16}.
It is beyond the scope of the present letter to undertake
an extensive search for the other parameters,
however, we can attest to the following.
Our results are insensitive to the value 
of the temperature, $T_{bb}$ , of the blackbody photons.
The case is similar for the parameter,
$z_0$, which denotes the distance of the base of the jet from the 
black hole.  This, however, will not be the case if general relativistic
effects are taken into account.
The width $z_1 - z_0$ of the acceleration
zone is not a crucial parameter either.  In contrast,
our results are sensitive to ${v_0}$ and 
$v_{\perp}$ and their values cannot stray too far from the 
ones that we used in this study. 
As $v_0$ decreases, the spectra become less anisotropic and, therefore,
the variation of $\Gamma$ with $\cos \theta$ is weaker. The opposite
happens when $v_0$ increases.
A small variation of $v_{\perp}$, on the other hand, can be accommodated 
by a corresponding variation of $\tau_{\parallel}$.
For example, a 10\% increase in $v_{\perp}$ 
can be accommodated by a similar decrease in $\tau_{\parallel}$.

The results presented in this work, as well as the 
results from our previous works
\citep{reig03,giannios04,giannios05,kylafis08,reig15,kylafis18,reig18,reig19}, 
show that 
Comptonization in the jet is a major source of the hard X-ray photons emitted by BHXBs. 
In the quiescent, hard, hard-intermediate, and soft-intermediate states,
a compact jet is fed from the corona.
Thus, a significant fraction 
of the soft photons that enter the corona to be Comptonized
cannot escape without entering the jet and also being scattered there. 

In addition to the explanation of the spectral variations in type-B QPOs,  our
jet model can  explain several observations and a number of correlations.  For
example, it can explain the correlation  between $\Gamma$ and time lags in GX
339-4 \citep{kylafis18}.  Such a correlation may actually be omnipresent in
BHXRBs  \citep{reig18}. This result cannot be explained by the propagating
fluctuations model \citep{lyubarskii97,kotov01,arevalo06,rapisarda17}, which
also provides a physical description of the time lags in BHXRBs.  It is possible
that this mechanism may take place in accretion flows, because fluctuations are
inevitable, but  no correlation between  $\Gamma$ and time lags would be
expected in this case because the timescale of propagating fluctuations is
determined by the properties of the accretion flow, while $\Gamma$ is determined
by the  optical depth and the temperature of the corona. On the other hand, time
lags due to Comptonization seem unavoidable, when there is a jet, because the
jet is situated above and below the corona. Finally, we note that our model not
only explains  existing correlations, but it also predicts new ones. Our jet
model predicts that a correlation between the break frequency  $\nu_{br}$ in
the  radio spectrum of GX 339-4 and the photon index $\Gamma$,  should be seen
during the rising part of a future outburst \citep{kylafis18}.

It seems appropriate to make some remarks regarding
similarities and dissimilarities between type-C and type-B QPOs.
In both, precession seems to play a central role, yet it
is widely accepted that two different mechanisms are involved in
these two types of QPOs \citep{motta11}.

For type-B QPOs, we have shown that they can be explained quantitatively with 
Comptonization in a precessing jet. The precessing jet is fed from a  precessing
corona (hot inner flow) and we note that such a precessing jet has already been
seen in  general relativistic magnetohydrodynamic simulations  \citep{liska18}.
Comptonization in the corona   can precede Comptonization in the jet, without
any significant  effect on the observed spectrum.   On the other hand, type-C
QPOs are nicely explained in a quantitative sense by Comptonization in a
precessing corona  \citep{ingram15,ingram16,ingram17,you18}, without any
involvement of the jet, which is definitely present.

Three questions naturally arise regarding
1) whether the jet is precessing when type-C QPOs are observed;  
2) whether Comptonization is taking place in the jet when we see type-C QPOs;
3) why we do not see the precessing corona in type-B QPOs.

The answer to the first question is 
probably yes, but unlike the type-B QPOs, for which the jet is narrow
(see below),
in type-C QPOs the jet is wide because the truncation radius is 
relatively large in the hard and hard-intermediate states. Thus,
even if the bottom part of the jet is precessing, there is hardly any 
variation in its observed emission. 

With regard to the second question, in type-C QPOs, the variability
at the QPO frequency comes directly
from the precessing corona.  Any QPO signal that enters the 
jet gets washed out because the light-travel time in the jet is much larger
than the period of the QPO.  Thus, Comptonization in the jet can happen, and
it almost certainly does, but 
the QPO signal comes directly from the corona. It may be that the ratio of 
the X-rays that are emitted by the corona and are not scattered by the jet 
over the X-rays emitted after Comptonization in the jet is constant, 
but the unscattered X-ray photons (in absolute number) from the corona, 
when it is large in size, is large enough for the QPO signal to be detected.  

In type-B QPOs on the other hand, the jet is narrow because the corona that
feeds it is very small.  This is because the truncation radius is close to 
the inner stable circular orbit in the soft intermediate state.  
Thus, the answer to the third question is that 
the QPO signal from the precessing 
corona is undetectably small and the variability in type-B QPOs comes
entirely from the precessing jet. 
We note here that the jet does not need to precess
as a solid body.  
We have found from our Monte Carlo code that the
upscattered photons reach, in their random walk in the jet,
a maximum height equal to a few times its radius, $R_0$. 
Thus, only the bottom part of the jet needs to precess as a solid body.

In summary, the hard X-rays come from Comptonization in the corona 
and the jet. In type-C QPOs, we are observing the precession of the corona,
whereas in type-B QPOs, we are observing the precession of the jet. 

\section{Conclusion}

In conclusion, we are able to quantitatively explain 
 the periodic variation of the 
photon index $\Gamma,$ observed by \citet{stevens16} in
GX 339-4, as coming from a precessing jet.  The values of the parameters 
that we use are reasonable for BHXRBs in the soft-intermediate state.

\begin{acknowledgements}
We thank the anonymous referee for asking critical questions, which helped
us in the qualitative discussion of our results.
We also thank Phil Uttley for a useful exchange of e-mails regarding 
type-C QPOs.
\end{acknowledgements}

\bibliographystyle{aa}
\bibliography{../../bhb} 

\end{document}